\documentclass[aps,prl,reprint]{revtex4-1}
\usepackage{graphicx}
\usepackage{dcolumn}
\usepackage{bm}
\usepackage{amssymb}
\usepackage{amsmath}
\usepackage{epsf}
\usepackage{color}
\begin{document}
\title{Magnetic structure and properties of a vanthoffite mineral $Na_6Mn(SO_4)_4$}

\author{Ajana Dutta,$^1$ Diptikanta Swain,$^2$ A. K. Bera,$^{3,4}$ Rajamani Raghunathan,$^5$
Debakanta Samal,$^{4,6}$ S. M. Yusuf,$^{3,4}$ S. Ramasesha,$^1$ and T. N. Guru Row$^1$}


\affiliation{$^1$Solid State and Structural Chemistry Unit, Indian Institute of Science, Bengaluru 560012, India}
\affiliation{$^2$Institute of Chemical Technology-Indian Oil Odisha Campus, Bhubaneswar 751013, India}
\affiliation{$^3$Solid State Physics Division, Bhabha Atomic Research Centre, Mumbai 40085, India}
\affiliation{$^4$Homi Bhabha National Institute, Anushaktinagar, Mumbai 400094, India}
\affiliation{$^5$UGC-DAE Consortium for Scientific Research, Indore 452001, India}
\affiliation{$^6$Institute of Physics, Bhubaneswar 751005, India}

\begin{abstract}
A detailed analysis of the magnetic properties of a vanthoffite type 
mineral $Na_6Mn(SO_4)_4$ based on dc magnetization, low temperature neutron 
powder diffraction and theoretical calculations is reported. The mineral 
crystallizes in a monoclinic system with space group $P2_1/c$, where 
$MnO_6$ octahedra are linked via $SO_4$ tetrahedra. This gives rise to super-exchange 
interaction between two $Mn^{2+}$ ions mediated by two nonmagnetic bridging 
anions and leads to an antiferromagnetic ordering below 3 K. 
The magnetic structure derived from neutron powder diffraction at 1.7 K depicts 
an antiferromagnetic spin arrangement in the $bc$ plane of the crystal. 
The magnetic properties are modelled by numerical calculations using 
exact diagonalization technique, which fits the experimental results  and
	provides antiferromagnetic ground state of $Na_6Mn(SO_4)_4$.
\end{abstract}
\maketitle
\section{Introduction} 
Over the last two decades, design of polyanionic materials 
[$(XO_4)_n$ with $X$ = S, P, As, V, Si, Mo or W] has attracted significant 
attention due to their adaptability towards various potential 
applications. For example, the discovery and commercialization 
of $LiFePO_4$ \cite{Padhi-JES-1997,Wang-EES-2015} significantly shed light on the  use of 
insertion materials in battery research with 
other polyanionic units \cite{Masquelier-CR-2013,Melot-ACR-2013,Gong-EES-2011,Rousse-CM-2014}. In this context, it may be noted 
that many naturally occurring minerals with a variety of polyanionic units offer 
a treasure trove of materials with associated tunable properties. 
Moreover, the presence of $3d$ transition metals in the chemical 
composition of such materials will open up the possibility of 
synthesizing solids with interesting magnetic behaviour. 
Several electrode materials inspired by the naturally occurring 
minerals have been investigated leading to the discovery of 
interesting magnetic properties in these materials \cite{Reynaud-IC-2013,
Reynaud-PRB-2014,Sun-CM-2015,Rousse-PRB-2017,Lander-IC-2016,Furrer-PRB-2020}. The coupling between 
magnetic and electrical properties in $3d$ metal based polyanionic compounds results in  
magnetoelectric effect \cite{Reynaud-PRB-2014,Lander-IC-2016,Petersen-PRB-2012,Li-PRB-2009,
Dell-EP-1965,Fogh-PRB-2017,Lander-JMCA-2015,Scaramucci-PRL-2012,Kornev-PRB-2000,
Fiebig-JPDAP-2005,Aken-Nature-2007,Bousquet-PRL-2011}, which has been successfully utilized 
to design various multiferroic materials \cite{Rivera-EPJB-2009,Rousse-PRB-2013,Yanda-ACPM-2020}. Specifically, the 
compound $TbPO_4$ \cite{Rado-PRB-1984,Bluck-JPC-1988} displays intrinsic bulk magneto-electric effect.
A series of polyanionic phosphates $LiMPO_4$  ($M$ = Mn, Co, Ni or Fe) have 
aroused special interest to evaluate the associated magnetic behaviour in these 
minerals \cite{Kornev-PRB-2000,Vaknin-PRB-2002,Rousse-CM-2003,Vaknin-PRL-2004,Gnewuch-IC-2020,Fogh-PRB-2019}.

The origin of magnetic interactions in these 
transition metal oxides, sulfates, phosphates and arsenates are 
governed by the overlap between 3d orbitals of the transition 
metal and 2p orbitals of the oxygen atom. Usually, in super 
exchange interactions, two magnetic metal centers are bridged via a single 
electronegative anion, like oxygen (M-O-M). However, in this material two 
metal centers interact via two oxygen atoms (M-O-O-M) and the magnetic 
interactions are hence weaker. A set of semi-empirical rules referred to 
as Goodenough-Kanamori-Anderson rules which these systems follow, 
are well described in the literature \cite{Anderson-PR-1959,Kanamori-JPCS-1959,Goodenough-PR-1955,Goodenough-JPCS-1958,Goodenough-PR-1961,Rousse-CM-2001}.

Many polyanionic compounds have been largely studied for their 
structural diversity where both types of interactions (M-O-M or M-O-O-M) 
are possible when changing the $3d$ transition metal as well 
as polyanions (e.g. $PO_4$, $SO_4$, $AsO_4$, $VO_4$ etc). For example, the 
magnetic structure of anhydrous $FeSO_4$ and $NiSO_4$ have antiferromagnetic 
sheets with ferromagnetic coupling between the sheets whereas in the 
case of $CoSO_4$, only antiferromagnetic ordering exists within each sheet. 
But the magnetic structure of $CrVO_4$ has ferromagnetically ordered 
sheets that stack antiferromagnetically. However, in each case, 
magnetic coupling involves a long super exchange pathway 
between magnetic centers via nonmagnetic sulfate or vanadate tetrahedron \cite{Frazer-PR-1962}. 
Beside these class of materials, other families of electrode materials 
such as fluorosulfates \cite{Melot-CM-2011,Melot-IC-2011,Melot-PRB-2012}, phosphates \cite{Kornev-PRB-2000,Vaknin-PRB-2002,Rousse-CM-2003,Rousse-CM-2001,Rousse-SSS-2002,Rousse-APA-2002} and 
borates \cite{Tao-IC-2013} have been studied and these materials also exhibit magnetic 
ordering at low temperatures. In fluorosulfates magnetic exchange interaction 
between nearest-neighbour ions is mediated either through M-F-M link or 
through M-O-O-M interaction via the oxygen anions at the sulfate tetrahedral edge.

Materials designed for potential battery electrode applications are 
also recognized as model compounds for their intriguing magnetic property; 
examples are marinate phases $Li_2M(SO_4)_2$ ($M$ = Mn, Fe or Co) and 
$LiFe(SO_4)_2$ \cite{Reynaud-IC-2013}. At low temperatures, these compounds show antiferromagnetic 
ordering due to a specific arrangement of transition metal octahedra ($MO_6$) 
and sulphate tetrahedra ($SO_4$). This particular structural arrangement 
solely enables the M-O-O-M exchange pathway between transition metal ions. 
Another interesting example in this series is the orthorhombic 
$Li_2Ni(SO_4)_2$ reported by Reynaud et al \cite{Reynaud-PRB-2014}. It has a particular 
arrangement of isolated $NiO_6$ octahedra which are interconnected via $SO_4$
tetrahedral units. As a result of exchange interaction between the 
$3d$ transition metal cations via two bridging ions, this 
phase is antiferromagnetic with a $T_N$= 28 K. Similar long-range 
antiferromagnetic ordering is also observed with 
isostructural orthorhombic $Li_2M(SO_4)_2$  ($M$ = Mn, Fe or Co) phase \cite{Lander-IC-2016}.

It is of interest to note that, Vanthoffite minerals occur 
in nature as oceanic salt deposits \cite{Keester-ACSB-1977,Dutta-IC-2020}. We have shown earlier 
that the crystal structure of $Na_6Mn(SO_4)_4$(a Vanthoffite mineral) is 
built from an alternating corner-sharing of $SO_4$ tetrahedra and transition metal 
octahedra $MnO_6$ resulting in an infinite two-dimensional framework in the $bc$ plane \cite{Sharma-IC-2017}.
Such specific connectivity suggests the possibility of long exchange pathway between 
two $Mn^{2+}$ centers via two oxygen atoms (Mn-O-O-Mn), which might lead to 
magnetic interaction akin to several other examples reported in 
the literature \cite{Reynaud-IC-2013,Reynaud-PRB-2014,Lander-IC-2016}. 
In this article, we investigate the magnetic 
structure of $Na_6Mn(SO_4)_4$ using variable temperature neutron diffraction.
 Besides, we have carried out exact diagonalization calculations of the model
 Hamiltonian to shed light on the magnetic properties and magnetic structure.

\section{Experimental Method}
Single crystals of $Na_6Mn(SO_4)_4$ were grown by slow 
evaporation at 80$^\circ$C from an aqueous solution containing 3:1 
stoichiometric molar ratio of $Na_2SO_4$ (Sigma-Aldrich, 99.99\%) 
and $MnSO_4.H_2O$ (Sigma-Aldrich, 99.99\%) as described in the 
earlier publication \cite{Sharma-IC-2017}. Colourless block-shaped crystals were obtained after 
15 days. The single crystal x-ray diffraction of the as grown crystal was 
carried out on an Oxford Xcalibur(Mova) diffractometer equipped 
with an EOS CCD detector and a microfocus sealed tube 
using MoK$\alpha$ X-radiation ($\lambda$ = 0.71073 \AA; 50 kV and 0.8 mA) and 
the structural parameters agree with the earlier report \cite{Sharma-IC-2017}.
Single crystals were crushed to form bulk polycrystalline powder for 
further characterization. Room temperature PXRD data was recorded on 
a PANalytical X'Pert PRO diffractometer using Cu K$\alpha$  
range of 8-60$^\circ$ using a step size of 0.013$^\circ$. X’Pert High 
Score Plus (version 4.8) \cite{Degen-PD-2014} was used to analyze the pattern and 
profile fitting refinements were carried out using the room temperature unit cell 
parameters of  $Na_6Mn(SO_4)_4$ \cite{Sharma-IC-2017} in JANA2006 \cite{Petricek-ZKM-2014}. 
Profile parameters such as 
GU, GV, GW, LX, and LY were refined using Pseudo-Voigt function. 
Neutron diffraction patterns over a wide $Q$-range 
($4\pi sin\theta/\lambda=0.3-9.5$\AA$^{-1}$ where $2\theta$ and $\lambda$ 
are the scattering angle and 
wavelength of the incident neutron beam, respectively) were recorded over 
1.7-300 K by using the powder diffractometer PD-II ($\lambda$ = 1.2443 \AA) 
at Dhruva reactor, Trombay, INDIA \cite{Siruguri-NN-2000}. For the neutron 
diffraction measurements, the powder sample was filled in a vanadium can of 
diameter 6 mm. All the low-temperature measurements were performed by using 
a closed-cycle helium refrigerator. The neutron diffraction patterns were analyzed 
by Rietveld refinement method using the FULLPROF suite program \cite{Carvajal-PB,Bera-JPCC-2020,Bera-MRE-2015,Bera-SSS-2013,Saha-SSC-2011}.
Temperature and magnetic field-dependent dc-susceptibility measurements were 
probed with a commercial vibrating sample magnetometer (Cryogenic Co. Ltd., UK). 
The temperature-dependent magnetization curves [$M$ vs. $T$] were recorded 
in the warming cycles over the temperature range of 2-300 K in 
both zero-field-cooled (ZFC) and field-cooled (FC) conditions. Isothermal magnetization 
curve was measured at 2 K in the increasing and decreasing field cycles up to 90 kOe.


\begin{figure}[t]
\centering
\includegraphics[scale=0.3]{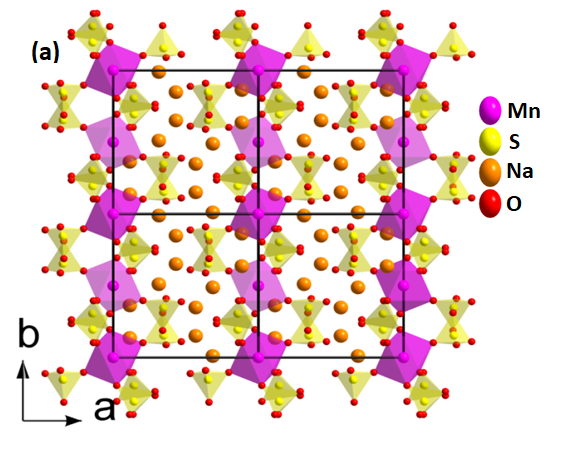}
\includegraphics[scale=0.4]{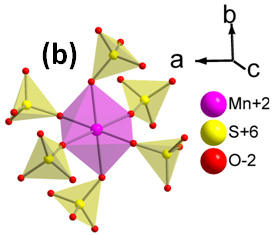}\\
\includegraphics[scale=0.4]{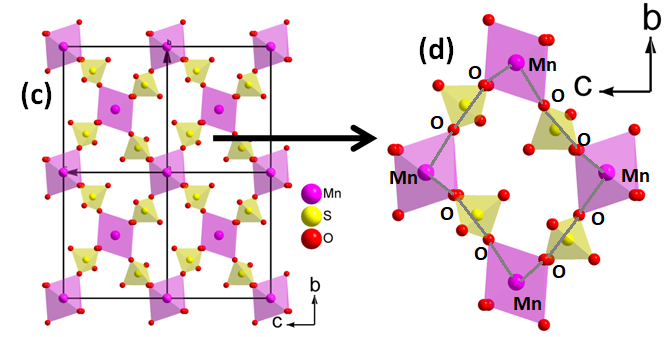}\\
\includegraphics[scale=0.3]{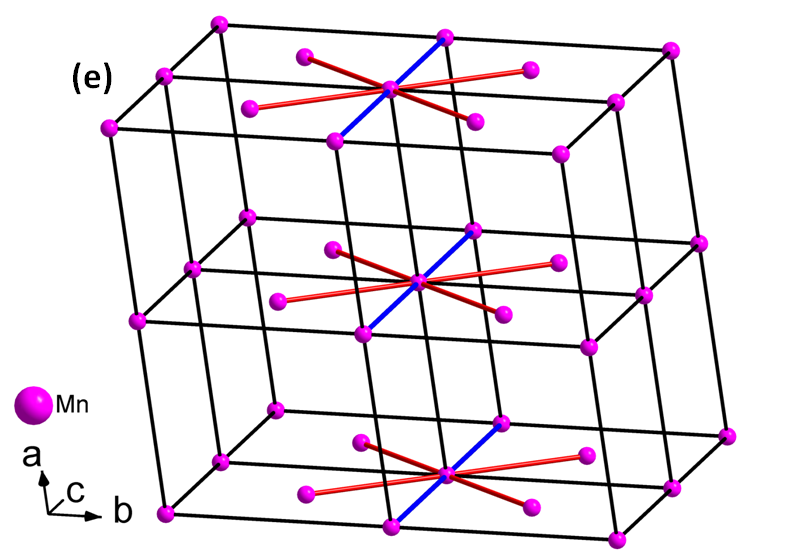}
\caption{\label{figure1} (a) Packing diagram of $Na_6Mn(SO_4)_4$ viewed down the $c$-axis,
(b) Highlighting the connectivity between $MnO_6$ and $SO_4$ tetrahedra,
(c) Packing diagram viewed down $a$-axis, sodium atoms are removed for
the clarity of the picture, (d) Mn-O-O-Mn interaction pathway through
$MnO_6$ octahedra and $SO_4$ tetrahedra, and
(e) Arrangement of Mn sublattice in the structure showing
the nearest neighbours (red bonds) and next-nearest neighbours (blue bonds). }
\end{figure}


\begin{figure}[t]
\centering
\includegraphics[scale=0.3]{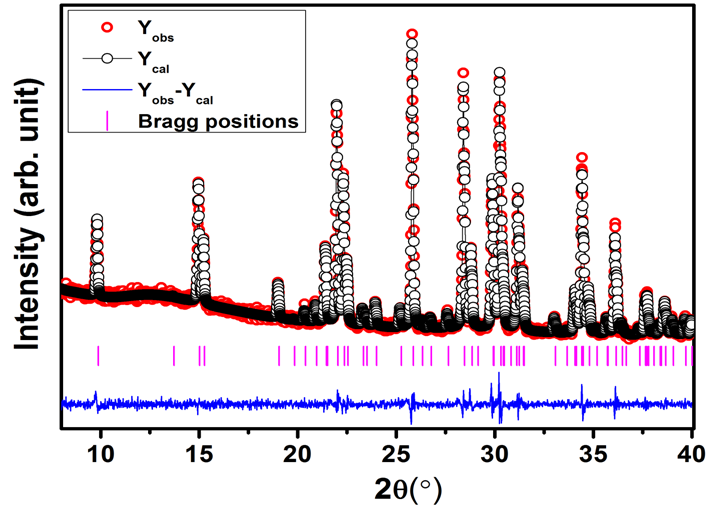}
\caption{\label{figure2}Le Bail profile refinement of $Na_6Mn(SO_4)_4$ at room temperature.}
\end{figure}

\section {RESULTS AND DISCUSSION}
$Na_6Mn(SO_4)_4$ belongs to a monoclinic system, space group $P2_1/c$ with Z = 2 as determined 
from single crystal X-ray diffraction for the present work and agrees 
well with the earlier report by our group \cite{Sharma-IC-2017}. The fractional coordinates of 
all atoms and the bond lengths and angles for $MnO_6$ octahedra are given in Table \ref{table1} and \ref{table2}. 
The asymmetric unit contains half the formula unit, where Mn atom 
is in a special position (Wyckoff position 2$a$, local site symmetry -1) along 
with three sodium atoms and two sulfate units in general position 
(Wyckoff position 4$e$, local site symmetry 1) [Table \ref{table1}]. Mn atom forms 
$MnO_6$ octahedra with symmetrically related oxygen atoms and connected 
to $SO_4$ tetrahedra in a ``pinwheel  pattern''. (Figure \ref{figure1}b) \cite{Brian-JEPM-1973}.


\begin{table*}[t]
\caption{\label{table1}Crystallographic details and fractional atomic coordinates for $Na_6Mn(SO_4)_4$}
\begin{tabular}{cccccccc}
\hline
\hline
\multicolumn{8}{c} {Empirical formula = $Na_6Mn(SO_4)_4$,  Formula weight (g/mol)=577.12, Space Group = $P2_1/c$,} \\
\multicolumn{8}{c} {$a$= 9.7131(13) \AA,  $b$ = 9.2926(11) \AA,  $c$ = 8.2609(12) \AA,  $\beta$ =112.988(7)$^\circ$,}\\
\multicolumn{8}{c} {$V$ = 686.42(16) \AA$^3$, $R_{obs}[I>2\sigma (I)]$ = 0.0183, $wR_{obs}[I>2\sigma (I)]$ =0.0541 } \\
\hline
\hspace{0.05in} {\bf Atom}\hspace{0.05in}  & \hspace{0.05in} {\bf Wickoff
	position}\hspace{0.05in}   & \hspace{0.05in} {\bf Occupancy}
	\hspace{0.05in}  & \hspace{0.15in} {$\mathbf x/a$}\hspace{0.15in}   &
	\hspace{0.15in} {$\mathbf y/b$}\hspace{0.15in}  & \hspace{0.15in}
	{$\mathbf z/c$}\hspace{0.15in}   &\hspace{0.15in}
	$\mathbf{u_{iso}(\AA^2)}$\hspace{0.15in}   & \hspace{0.15in} {\bf BVS}\hspace{0.15in}   \\
\hline
Mn1  &  2$a$  &   0.5  &  0.000000         & 0.000000      & 0.000000      & 0.01093(11)   & 1.927\\
Na1  &  4$e$  &   1    &  0.11489(9)       & 0.36352(8)    & 0.18562(11)   & 0.02590(19)   & 1.079\\
Na2  &  4$e$ &   1    &  0.31418(8)       & -0.01160(7)   & 0.46824(9)    & 0.01645(17)   & 1.105\\
Na3  &  4$e$ &   1    &  0.43369(9)       & -0.15142(8)   & 0.07766(10)   & 0.02486(19)   & 1.060\\
S1   &  4$e$ &   1    &  0.34511(4)       & 0.15332(4)    & 0.16518(5)    & 0.01101(12)   & 6.028\\
S2   &  4$e$ &   1    &  0.14234(4)       & -0.30579(4)   & 0.21800(5)    & 0.01055(12)   & 6.060\\
O1   &  4$e$ &   1    &  0.20375(13)      & 0.10249(13)   & 0.17787(15)   & 0.0159(3)     & 2.038\\
O2   &  4$e$ &   1    &  0.33934(14)      & 0.31084(13)   & 0.15062(16)   & 0.0173(3)     & 2.134\\
O3   &  4$e$ &   1    &  0.36223(15)      & 0.08889(14)   & 0.01351(17)   & 0.0222(3)     & 2.043\\
O4   &  4$e$ &   1    &  0.46806(13)      & 0.11001(14)   & 0.32974(17)   & 0.0201(3)     & 2.110\\
O5   &  4$e$ &   1    &  0.02432(14)      & -0.19349(13)  & 0.15358(17)   & 0.0176(3)     & 2.046\\
O6   &  4$e$ &   1    &  0.28843(14)      & -0.23682(14)  & 0.30234(18)   & 0.0218(3)     & 1.979\\
O7   &  4$e$ &   1    &  0.13626(14)      & -0.39575(13)  & 0.06987(16)   & 0.0196(3)     & 2.061\\
O8   &  4$e$ &   1    &  0.11064(16)      & -0.39682(14)  & 0.34471(18)   & 0.0229(3)     & 1.983\\
\hline
\hline
\end{tabular}
\end{table*}

\begin{table*}[t]
\caption{\label{table2} Bond lengths and bond angles for $Na_6Mn(SO_4)_4$}
\begin{tabular}{cccccc}
\hline
\hline
\hspace{0.2in} {\bf Bond}\hspace{0.2in}  & \hspace{0.2in} {\bf Length (\AA)}\hspace{0.2in}   & \hspace{0.2in} {\bf Bond} \hspace{0.2in}  & \hspace{0.2in} {\bf Angle ($^\circ$) } \hspace{0.2in}  & \hspace{0.2in}  {\bf Bond}\hspace{0.2in}  & \hspace{0.2in} {\bf Angle ($^\circ$)}\hspace{0.2in}   \\
\hline
Mn1-O5x2   &    2.1597(12)   &  O5-Mn1-O1x2    &  96.08(5)   &  O5-Mn1-O8x2   & 90.78(5) \\
Mn1-O1x2   &    2.1706(12)   &  O5-Mn1-O1x2    &  83.92(5)   &  O1-Mn1-O8x2   & 86.42(5) \\
Mn1-O8x2   &    2.1901(13)   &  O5-Mn1-O8x2    &  89.22(5)   &  O1-Mn1-O8x2   & 93.58(5) \\
\hline
\hline
\end{tabular}
\end{table*}

The Mn-O bond lengths in $MnO_6$ octahedra varies between 2.1597 (12) 
to 2.1901(13) \AA (Table \ref{table2}) where the bond length distortion parameters and 
bond angle variance are calculated using formulas 
$\Delta d=(\frac{1}{6})\sum_n[\frac{d_n-d_{av}}{d_{av}}]^2$ and 
$\sigma^2_{oct}=\frac{1}{11}\sum_i[\alpha_i-90]^2$, ($d_n$ and $d_{av}$ are 
the individual and average Mn-O bond length and $\alpha_i$ are the individual O-Mn-O 
bond angles) \cite{Lu-JACS-2020,Baur-ACSB,Robinson-Science-1971}. It is to be noted that, $\Delta d$ and $\sigma_{oct}^2$ values 
for an ideal octahedron should be exactly zero. The bond length distortion 
parameter obtained ($\Delta d$=3.35$\times$10$^{-5}$) though indicates a quite 
symmetrical $MnO_6$ octahedra, the calculated bond angle variance of 18.32 
show a distorted $MnO_6$ octahedra. The bond valence sum for Mn 
atom is calculated to be around 1.927 using the Zachariasen formula 
$V_i=\sum_js_{ij}=\sum_je^{\frac{d_o-d_{ij}}{0.37}}$
and it is in good agreement with the expected valance of +2 \cite{Brown-ACB-1985}. 
These are isolated $MnO_6$ octahedra (pink) and are connected to 
$SO_4$ tetrahedra (yellow) via their oxygen vertices (Figure \ref{figure1}). 
Thus the structure presents an exchange pathway via two bridged oxygen 
atoms viz., Mn-O-O-Mn magnetic interaction where Mn-O-O-Mn 
dihedral angle is about 148$^\circ$ (Figure \ref{figure1}d). A similar long exchange 
pathway (M-O-O-M) is found in $Li_2M(SO_4)_2$, ($M$ = Ni, Co, Fe, Mn), where 
magnetism in the materials are explained based on this interaction \cite{Reynaud-IC-2013,Reynaud-PRB-2014,Lander-IC-2016}. 
The single crystals grown are further crushed to form the powder sample and 
the phase purity was checked using PXRD measurement. The PXRD profile 
refinement ($R_p$ = 3.43, $R_{wp}$ = 4.50 and $\chi^2$ = 1.01) 
at room temperature was carried out using the cell parameter and 
space group obtained from the single crystal XRD, where the close 
similarity between the observed and the calculated patterns suggests the purity 
of the desired compound, $Na_6Mn(SO_4)_4$ (Figure \ref{figure2}).

\subsection{Magnetic Measurements}


\begin{figure}[t]
\centering
\includegraphics[scale=0.07]{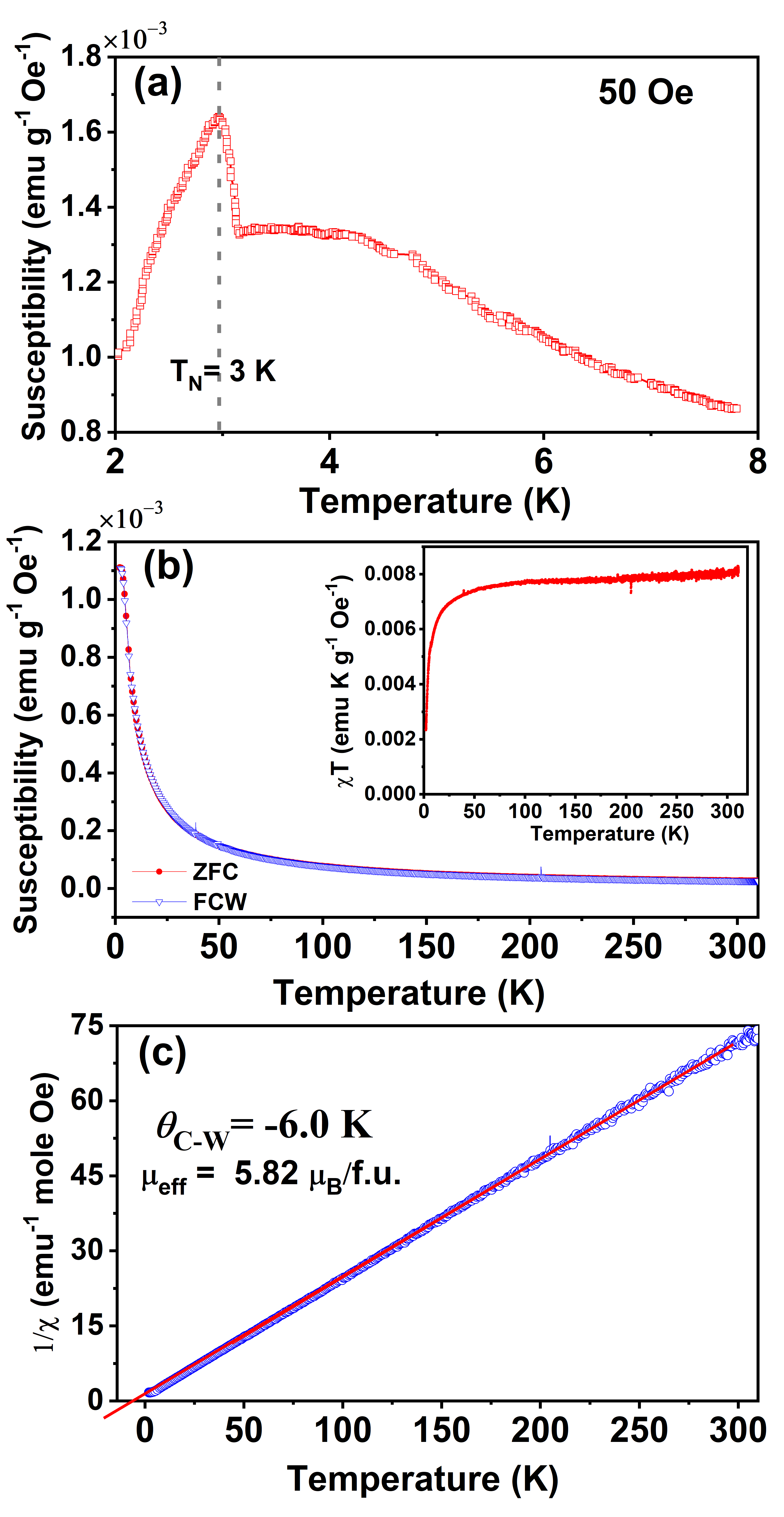}
\caption{\label{figure3} (a) Low temperature susceptibility curve
 measured under 50 Oe in  ZFC  mode, (b) The temperature-dependent
 susceptibility ($\chi$(T)) curves measured under 1000 Oe in the
 ZFC and FC modes. Inset shows $\chi T$ {\it vs.} $T$ plot in the ZFC mode, and
 (c) The inverse ZFC susceptibility as a function of temperature
 under 1000 Oe. The solid curve is a straight line fit to the measured data. }
\end{figure}


\begin{figure}[t]
\centering
\includegraphics[scale=0.3]{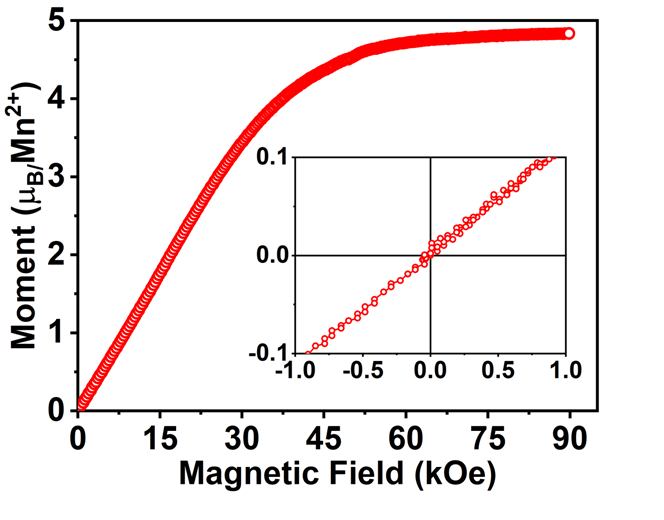}
\caption{\label{figure4} The isothermal magnetization measured
 at 2 K after cooling the sample in zero field. Inset shows
 a zoomed view in the low field region
 over $\pm$ 1 kOe and absence of hysteresis.}
\end{figure}

The zero-field-cooled susceptibility curve of $Na_6Mn(SO_4)_4$ 
measured under magnetic field of 50 Oe (Figure \ref{figure3}a) shows a peak 
revealing a transition to antiferromagntic (AFM) 
state below $T_N$ $\sim$ 3 K. The AFM ordering is confirmed by 
our zero field neutron diffraction study presented later. Figure \ref{figure3}b shows the temperature 
dependent susceptibility curve measured under 1000 Oe over the temperature range 
2-300 K. The  $\chi$ vs T curve in Figure \ref{figure3}b, however 
does not exhibit any distinct magnetic transition in contrast to the observation 
made under a weak applied field of 50 Oe in Figure \ref{figure3}a. It is to be 
noted that ZFC $\chi T$ vs T plot under 1000 Oe [inset of Figure \ref{figure3}b] 
yields a downturn below 20 K. This corroborates the existing antiferromagnetic 
interactions in $Na_6Mn(SO_4)_4$. We also notice that $\chi T$ value in 
the high temperature (paramagnetic) region increases slightly with temperature, 
contrary to the constant value expected in the paramagnetic region.  
The reason for the same could be attributed to additional contribution 
arising from the van Vleck paramagnetism ($\chi_{VV}$) which is discussed in the theoretical section.

The inverse ZFC susceptibility plot under 1000 Oe is shown in Figure \ref{figure3}c. 
The linear fit to the inverse susceptibility curve 
yields the Curie-Weiss temperature $\Theta_{CW}$ = -6.0 K and 
the effective paramagnetic moment $\mu_{eff}$ = 5.8 $\mu_B$/f.u. 
The observed value of the effective moment 5.8 $\mu_B$/f.u is 
in good agreement with the theoretically expected $\mu_{eff}$ 
(where $\mu_{eff}(s) = g\sqrt{s(s + 1)}$ with  g=2 (Lande g-factor))
value of 
5.92 $\mu_B$/Mn$^{+2}$, considering only spin moment. This result confirms +2 
oxidation state of the magnetic Mn ion (s=5/2) 
in $Na_6Mn(SO_4)_4$. The isothermal field dependent 
magnetization curve (Figure \ref{figure4}) measured at 2 K, shows a 
linear increase in the low field regime and then tends to 
show a change in slope above 35 kOe and a saturation above 65 kOe. However, we do not observe any 
opening of the hysteresis loop (inset in Figure \ref{figure4}) under 
field sweeping. The observation of negative Curie-Weiss temperature, 
downturn of $\chi T$ vs T, linear magnetization behaviour in the low 
field region and the absence of hysteresis altogether suggest 
an antiferromagnetic ground state of $Na_6Mn(SO_4)_4$.

\subsection{Neutron Diffraction}

\begin{figure}[t]
\centering
\includegraphics[scale=0.070]{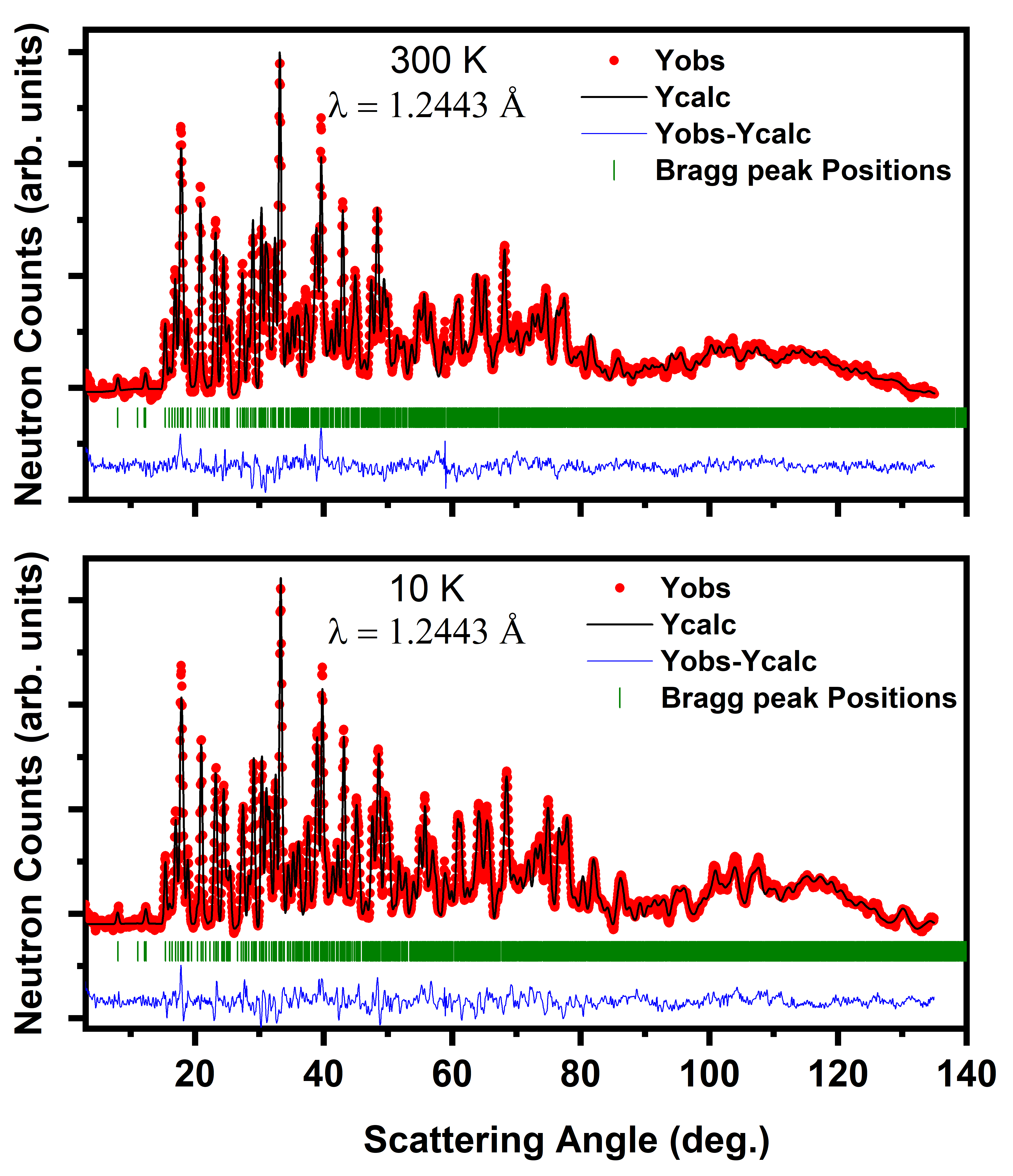}
\caption{\label{figure5} Rietveld refinement of Neutron diffraction patterns for
	$Na_6Mn(SO_4)_4$ measured at (top) 300 K and (bottom) 10 K.}
\end{figure}

In order to further investigate the magnetic 
ground state of the material, neutron diffraction 
data were collected on bulk powder sample over 
the temperature range 1.7-300 K. Preparation of phase pure 
compound in sufficient quantity to perform neutron diffraction is rather challenging. 
However, almost 5 gm of single crystals were grown in different batches 
and these crystals were crushed to form polycrystalline powdered sample. 
Phase purity of bulk amount of powdered sample was 
checked via laboratory PXRD. 
The Rietveld refined neutron diffraction patterns measured at 
300 K and 10 K are shown in Figure \ref{figure5}. The crystal structure 
for $Na_6Mn(SO_4)_4$ remains monoclinic with space group $P2_1/c$  
over the entire temperature range 1.7-300 K.


\begin{figure}[t]
\centering
\includegraphics[scale=0.3]{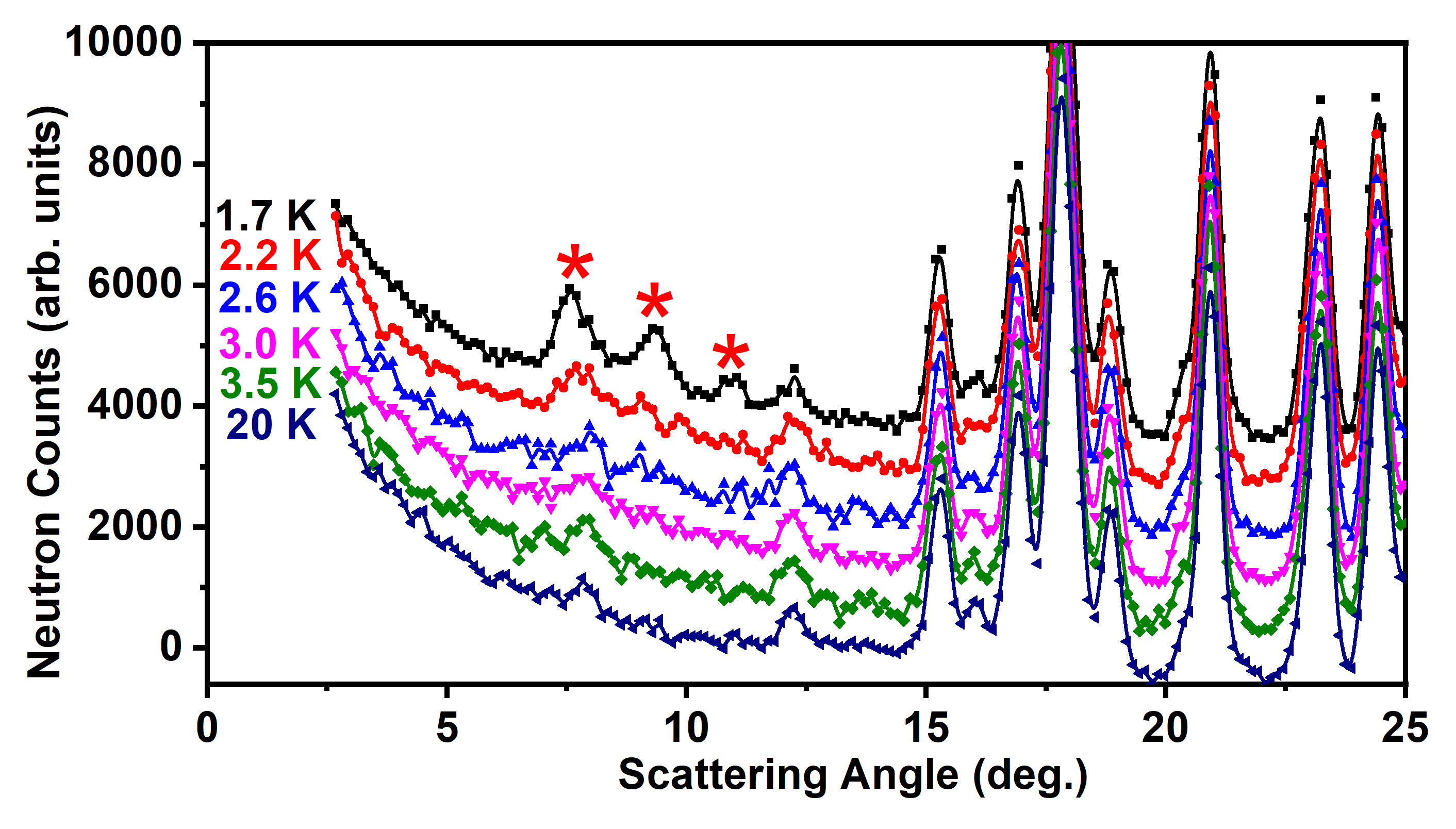}
\caption{\label{figure6} Low-temperature neutron diffraction patterns down to 1.7 K}
\end{figure}

In order to probe the long-range antiferromagnetic 
interaction in $Na_6Mn(SO_4)_4$, neutron diffraction data were 
collected down to 1.7 K (Figure \ref{figure6}). The appearance of additional magnetic 
Bragg peaks at $2\theta~\sim$  7.6$^\circ$, 9.4$^\circ$, and 11$^\circ$ 
(Marked with asterisks in Figure \ref{figure6}) below 3 K confirms a long-range 
antiferromagnetic ordering of the material.

All magnetic reflections observed for $Na_6Mn(SO_4)_4$ could be indexed 
with a propagation vector {\bf k} = (0,0,0) with respect to the same 
monoclinic unit cell as the nuclear structure. The symmetry-allowed 
magnetic structure is determined by a representation analysis, as 
applied for various kinds of spin systems \cite{Bera-PRB-2022,Suresh-PRB-2018,Bera-PRB-2017,Bera-PRB-2016}, 
using the program BASIREPS available with the FULLPROF 
program suite \cite{Carvajal-PB}. The results of the symmetry analysis reveal 
that there are four irreducible representations (IRs). 
Among the four IRs, the IR(1) or $\Gamma_1$ and IR(3) or $\Gamma_3$ are non-zero 
for the magnetic site of the present compound. Therefore, there are 
two possible symmetry allowed magnetic structures for $Na_6Mn(SO_4)_4$. 
Both the IRs $\Gamma_1$ and $\Gamma_3$ are one-dimensional.

\begin{table}[t]
\caption{\label{table3} Basis vectors of the magnetic sites of Mn with the propagation vector {\bf k} = (0 0 0) for  $Na_6Mn(SO_4)_4$. Only the real components of the basis vectors are presented. The two atoms of the non-primitive basis are defined according to Mn-1 (x, y, z):(0.5, 0, 0.5) and Mn-2 (-x, y+1/2, -z+1/2): (-0.5, 0.5, 0).}
\begin{tabular}{|c|c|c|c|}
\hline
\hspace{0.2in} {\bf IRs}\hspace{0.2in}  & \hspace{0.2in} \hspace{0.2in}   & \multicolumn{2}{c|} {\hspace{0.2in} {\bf Basis vectors} \hspace{0.2in}}  \\ \hline
            &            &       \multicolumn{2}{c|} {\hspace{0.2in} Site (2$b$) \hspace{0.2in} }\\\hline
            &            &       Mn-1    &     Mn-2    \\\hline
$\Gamma_1$  &  $\Psi_1$  &    100        &   -100      \\\hline
            &  $\Psi_2$  &    010        &    010      \\\hline
            &  $\Psi_3$  &    001        &    00-1     \\\hline
$\Gamma_2$  &  $\Psi_1$  &    100        &    100      \\\hline
            &  $\Psi_2$  &    010        &    0-10      \\\hline
            &  $\Psi_3$  &    001        &    001     \\
\hline
\end{tabular}
\end{table}


\begin{figure}[t]
\centering
\includegraphics[scale=0.3]{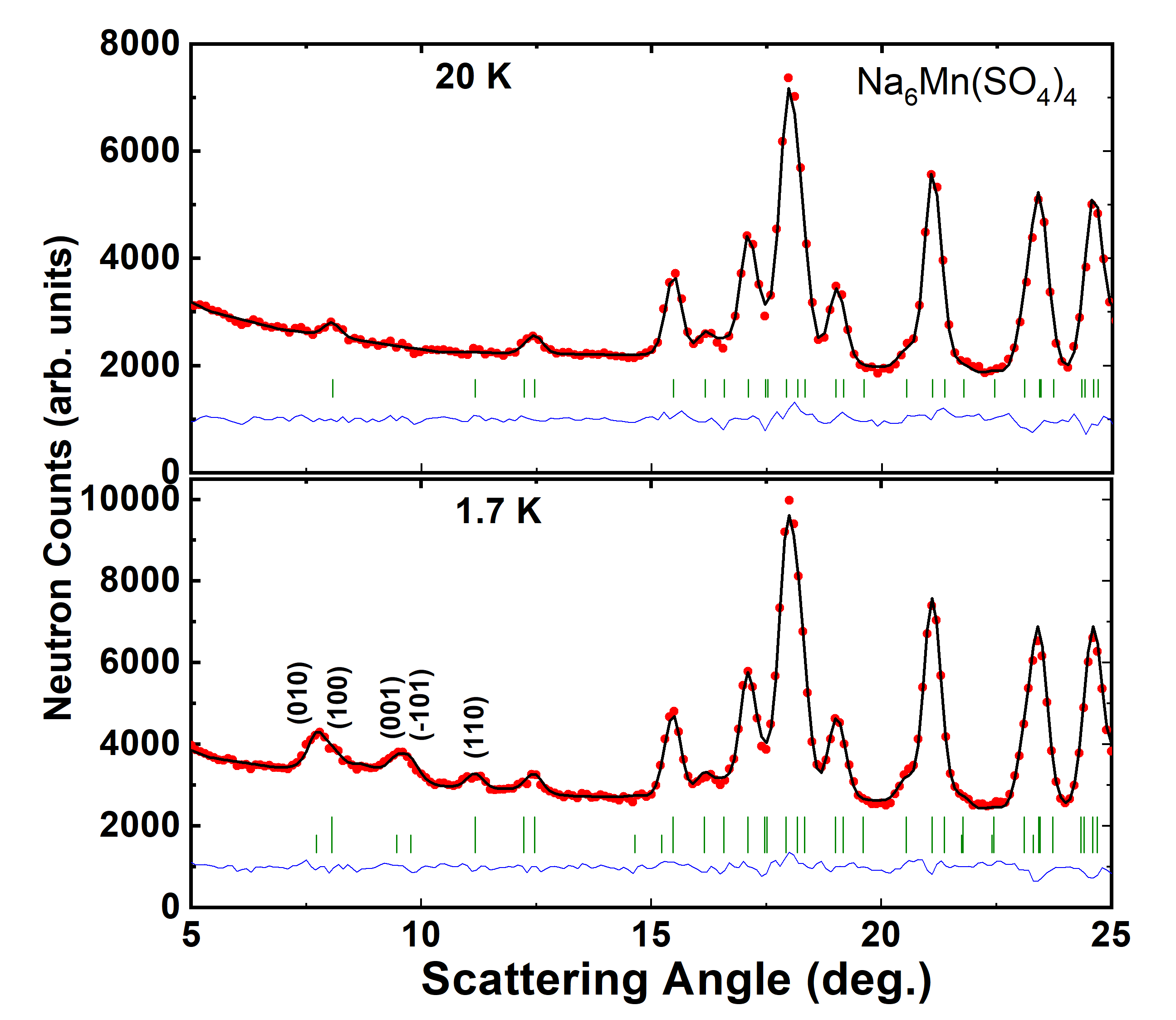}
\caption{\label{figure7} Experimentally observed (circles)
 and calculated (solid lines through the data points) neutron
 diffraction patterns for $Na_6Mn(SO_4)_4$ at
 (top) 20 K (paramagnetic state) and
 (bottom) 1.7 K (magnetically ordered state), respectively. The solid
 lines at the bottom of each panel represent the
 difference between observed and calculated patterns. The vertical bars
 indicate the positions of allowed nuclear
 and magnetic [the bottom panel] Bragg peaks. }
\end{figure}


\begin{figure}[t]
\centering
\includegraphics[scale=0.3]{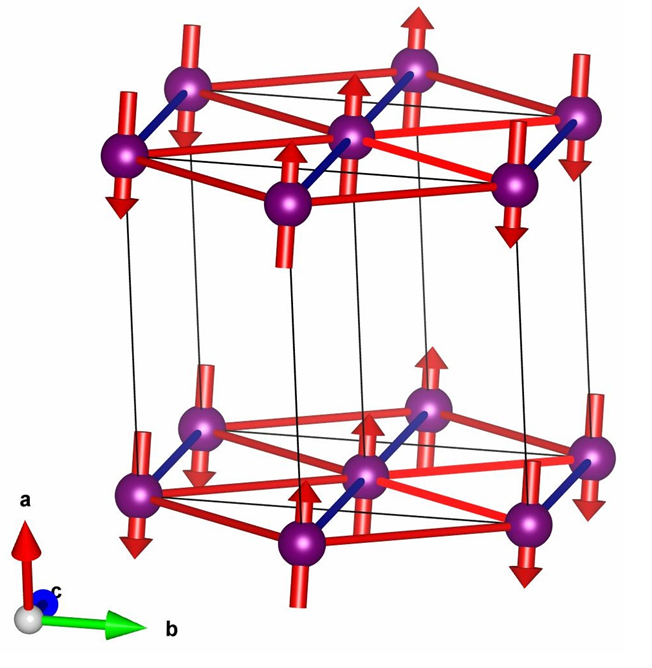}
\caption{\label{figure8} The magnetic structure of $Na_6Mn(SO_4)_4$.}
\end{figure}


\begin{figure}[t]
\centering
\includegraphics[scale=0.075]{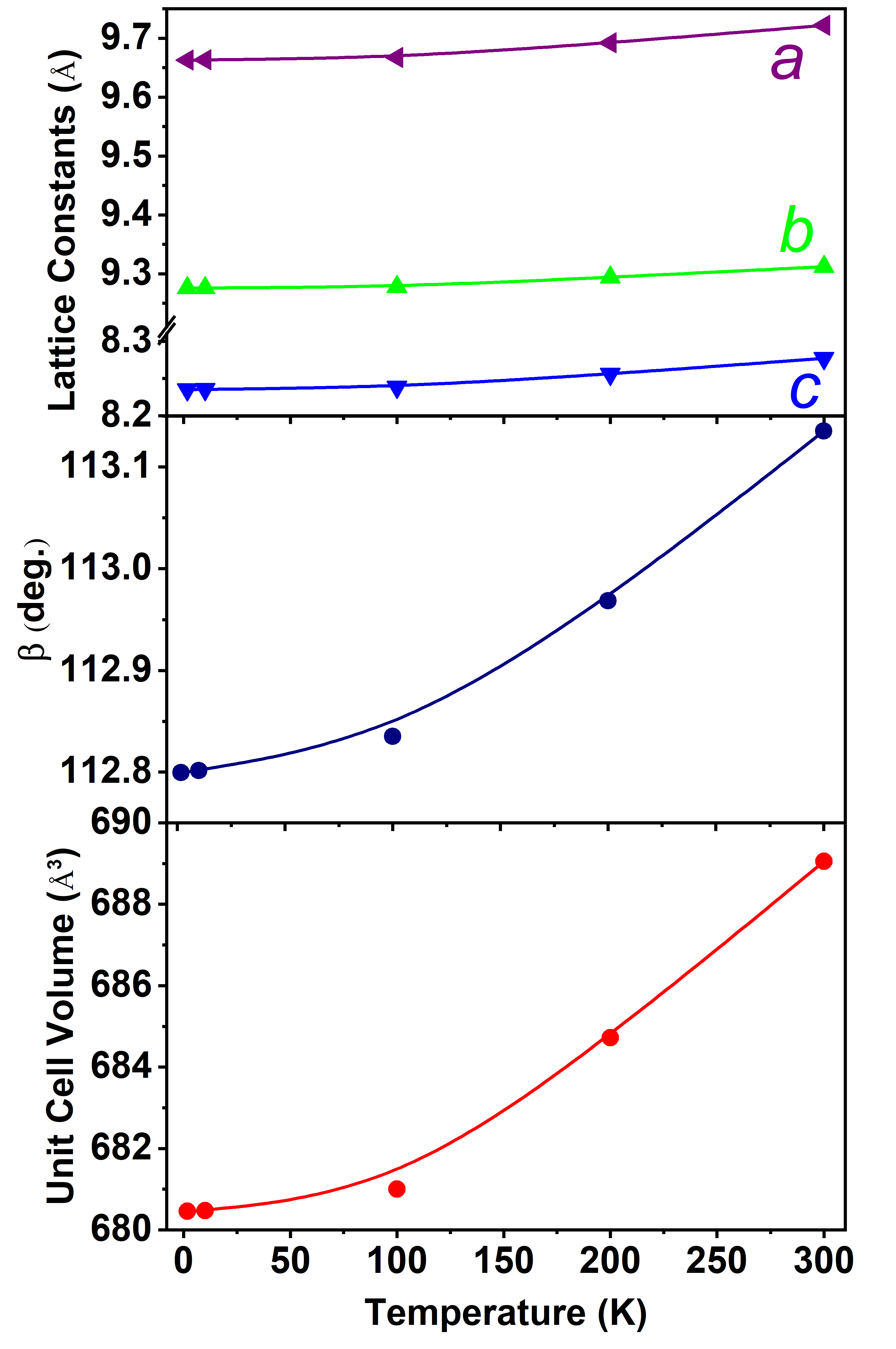}
\caption{\label{figure9} The temperature-dependent lattice parameters
 and unit cell volume of $Na_6Mn(SO_4)_4$ over the temperature range 1.7-300 K.}
\end{figure}

The magnetic representation $\Gamma_{mag}$ is composed as

\begin{equation}
\label{eqn_mag-rep}
  \Gamma_{mag}=3\Gamma_1+3\Gamma_3
\end{equation}

The basis vectors (the Fourier components of the magnetization) 
for these two IRs $\Gamma_1$ and $\Gamma_3$ for the magnetic site 
are given in Table \ref{table3}. The basis vectors are calculated using 
the projection operator technique implemented in the BASIREPS 
program \cite{Carvajal-PB,Carvajal-Fullprof}. Out of the $\Gamma_1$ and $\Gamma_3$, the best refinement of 
the magnetic diffraction pattern is obtained for 
the IR $\Gamma_1$. The refinement with the $\Gamma_1$ is shown in Figure \ref{figure7}. 
A good agreement is observed between observed and calculated pattern.

The corresponding magnetic structure is shown in Figure \ref{figure8}. 
The magnetic structure reveals antiferromagnetic chains of the Mn 
moments along the NN bond (red bonds) directions in the $bc$ plane, 
and such chains are coupled ferromagnetically along the NNN bond 
(blue bonds) directions in the $bc$ plane. 
Therefore, the magnetic structure within the $bc$ plane is 
a N\'eel type AFM. Such antiferromagnetic planes are stacked ferromagnetically 
along the $a$-axis (grey bonds). The magnetic structure is 
purely antiferromagnetic in nature without having any net magnetization per 
unit cell. The magnetic moments are lying in the $ac$ plane with 
moment components $m_a$ = 2.60(8) and $m_c$= 1.35(28) $\mu_B$ per magnetic site (Mn$^{2+}$) along 
the $a$ and $c$ axes respectively. The net ordered site 
moment of Mn ions (considering all the components) is 
found to be $M_{total}$ = 2.42 (3) $\mu_B$/Mn$^{2+}$ at 1.7 K. The magnetic moment 
is found to be strongly reduced from the theoretically expected value 
of 4 $\mu_B$/Mn$^{2+}$ ($\sim$ 80 $\%$ of the fully ordered moment of 5 $\mu_B$/Mn$^{2+}$) revealing the presence of a strong spin fluctuation 
at 1.7 K. The temperature variation of the lattice parameters and unit cell volume 
is shown in Figure \ref{figure9}. Change in slope at low temperature could 
be due to the interaction with magnetic spin and lattice.

\subsection{Theoretical study of $Na_6Mn(SO_4)_4$}

\begin{figure}[ht]
\centering
\includegraphics[scale=0.20]{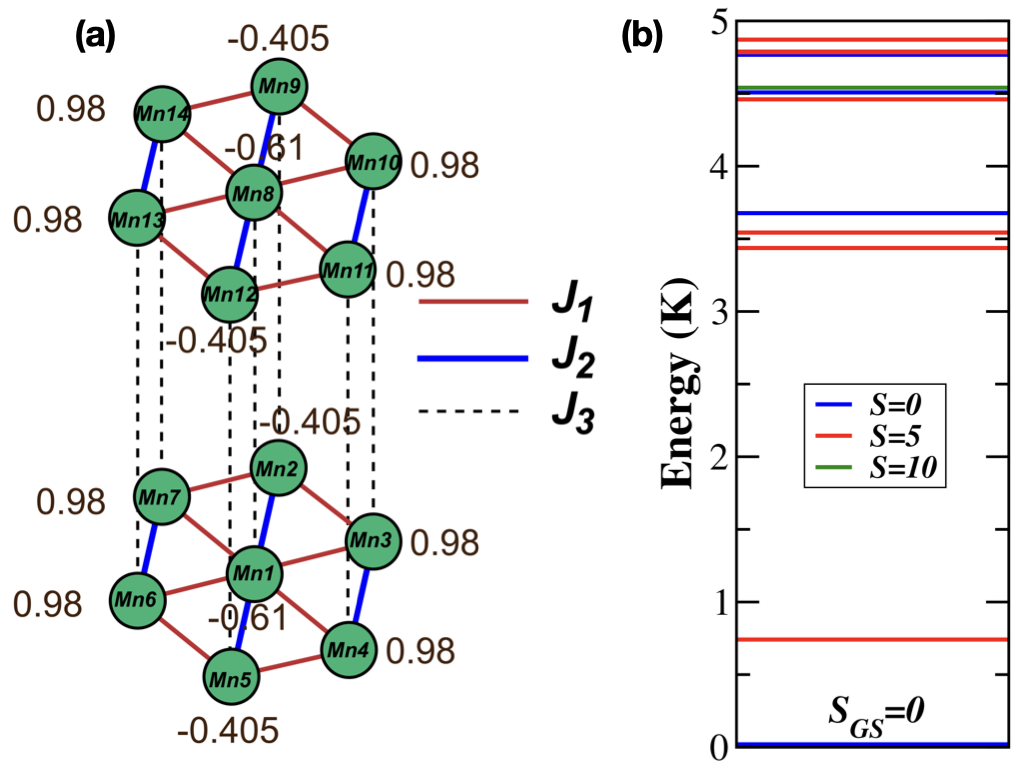}\\\vspace{0.2in}
\includegraphics[scale=0.35]{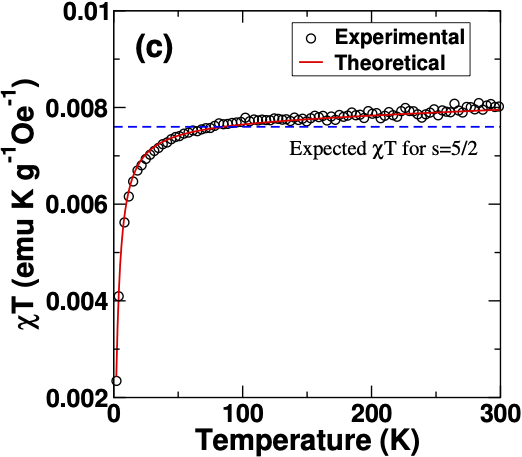}
\caption{\label{figure10} (a) Schematic of the
 magnetic exchange interactions in $Na_6Mn(SO_4)_4$.
 The spin densities of $S$=5, $M_S$=+5 state are also shown.
 The spin density at each Mn site is scaled by a 
 factor of 5 to convert from $s$=1/2 to $s$=5/2 site spin, 
 (b) Low-lying energy eigen states of our model in Kelvin along with the 
 total spin of the state,  
 and (c) Magnetic susceptibility ($\chi T$) as a function of temperature.
 Experimental points are shown by circles and calculated values fall
 on the red line. Blue line indicates the expected $\chi T$ value for free $s$=5/2 moments. }
\end{figure}

The refined X-ray diffraction data of $Na_6Mn(SO_4)_4$ (Figure \ref{figure1}) 
shows a primitive monoclinic crystal structure in which $Mn^{2+}$ 
ions are placed at each corner of the unit cell 
and an additional $Mn^{2+}$ ion is located at the 
face-center position in the $bc$ plane. A careful analysis of 
the structural information reveals that any $Mn^{2+}$ ion located at the 
corner of the unit cell is connected to four first nearest neighbours 
along the face diagonal in $bc$ plane and two second neighbours 
along the $c$-axis. This arrangement repeats along the $a$-axis, as shown 
in the Figure \ref{figure10}a. Heisenberg Hamiltonian is solved on the minimum cluster which adequately 
represents the crystal. This involves fourteen $Mn^{2+}$ ions at 
the vertices and at the centre of two hexagons parallel to each 
other, as shown in Figure \ref{figure10}a. The spin of each $Mn^{2+}$ ion is 5/2 
as the crystal field is weak. Exact diagonalization of the 
14 site s = 5/2 spin Heisenberg system is computationally prohibitive as the number of spin 
orientations (dimensionality of the Fock space) is 
more than 78 billion. Hence we have replaced the s = 5/2 
site spins by s = 1/2 site spins and have scaled the computed 
susceptibility by a factor of 11.67 which is the ratio of the 
square of the magnetic moments of a s = 5/2 ion and s = 1/2 ion. 
The Fock space dimension of the 14 spin-1/2 system is only 16,384. 
Furthermore, since z-component of the total spin, $S_z$ is conserved, we can factor the 
space into different $M_S$ sectors. Solving the eigen system for all the 
eigenvalues and eigenvectors is not compute intensive and 
affords exploring the parameter space of the exchange constants in the Hamiltonian 
on a fine grid.


The magnetic properties are modelled by employing the Heisenberg spin Hamiltonian,

\begin{eqnarray}
\label{eqn-model-H}
 \hat{H}_o & = &-J_1(\hat{\vec{s}}_1\cdot \hat{\vec{s}}_3 + \hat{\vec{s}}_1\cdot \hat{\vec{s}}_4 + \hat{\vec{s}}_1\cdot \hat{\vec{s}}_6 + \hat{\vec{s}}_1\cdot \hat{\vec{s}}_7 + \hat{\vec{s}}_2\cdot \hat{\vec{s}}_3 \nonumber\\
 && + \hat{\vec{s}}_4\cdot \hat{\vec{s}}_5 + \hat{\vec{s}}_5\cdot \hat{\vec{s}}_6 + \hat{\vec{s}}_2\cdot \hat{\vec{s}}_7 + \hat{\vec{s}}_{8}\cdot \hat{\vec{s}}_{10} + \hat{\vec{s}}_{8} \cdot \hat{\vec{s}}_{11} \nonumber\\ 
&& +  \hat{\vec{s}}_{8} \cdot \hat{\vec{s}}_{13} + \hat{\vec{s}}_{8} \cdot \hat{\vec{s}}_{14} + \hat{\vec{s}}_{9} \cdot \hat{\vec{s}}_{10} + \hat{\vec{s}}_{11} \cdot \hat{\vec{s}}_{12} +  \hat{\vec{s}}_{12} \cdot \hat{\vec{s}}_{13} \nonumber\\ 
&& + \hat{\vec{s}}_{9} \cdot \hat{\vec{s}}_{14}) - J_2 (\hat{\vec{s}}_1\cdot \hat{\vec{s}}_2 + \hat{\vec{s}}_{3} \cdot \hat{\vec{s}}_{4} + \hat{\vec{s}}_{1} \cdot \hat{\vec{s}}_{5} + \hat{\vec{s}}_{6} \cdot \hat{\vec{s}}_{7} \nonumber\\ 
&& + \hat{\vec{s}}_{8} \cdot \hat{\vec{s}}_{9} + \hat{\vec{s}}_{10} \cdot \hat{\vec{s}}_{11} + \hat{\vec{s}}_{8} \cdot \hat{\vec{s}}_{12} + \hat{\vec{s}}_{13} \cdot \hat{\vec{s}}_{14}) -J_3(\hat{\vec{s}}_{1} \cdot \hat{\vec{s}}_{8} \nonumber\\ 
&& + \hat{\vec{s}}_{2} \cdot \hat{\vec{s}}_{9} + \hat{\vec{s}}_{3} \cdot \hat{\vec{s}}_{10} + \hat{\vec{s}}_{4} \cdot \hat{\vec{s}}_{11} + \hat{\vec{s}}_{5} \cdot \hat{\vec{s}}_{12} + \hat{\vec{s}}_{6} \cdot \hat{\vec{s}}_{13} \nonumber\\ 
&& + \hat{\vec{s}}_{7} \cdot \hat{\vec{s}}_{14})
\end{eqnarray}

where, $J_1$, $J_2$ and $J_3$ are the strength of exchange 
interactions between first, second and third neighbours, respectively and 
$\hat{\vec{s}}$ are the site spin operators and the 
numbers in the subscript represent the site index as in 
Figure \ref{figure10}a. A positive or negative value of $J$ 
corresponds to a ferromagnetic or antiferromagnetic exchange interaction 
respectively. The three unique exchange parameters $J_1$, $J_2$ and $J_3$ 
are all antiferromagnetic and 
have their strengths that are exponentially dependent on the distance 
between ions; hence $|J_1| > |J_2| > |J_3|$. The exchange constants $J_2$ and $J_3$ 
are expressed as fractions of $J_1$, which is set to -1.0. 
We have taken the two exchange constants $J_2$ and $J_3$ 
as $J_2 = -e^{-\frac{r_2}{r_1}}$ and $J_3 = -e^{-\frac{r_3}{r_1}}$, where 
$r_1$, $r_2$ and $r_3$ are the first, second and third neighbour distances from the 
refined X-ray diffraction data. As the first neighbour 
Mn-O-O-Mn dihedral angle is about 148$^\circ$ (from X-ray structure), 
we take $J_1$ to be antiferromagnetic. 

The matrix of the spin Hamiltonian (eq. \ref{eqn-model-H}) was constructed 
using a basis with constant total $M_S$. The largest Hamiltonian 
matrix which is 3432 x 3432 corresponds to the $M_S=0$ sector. 
We obtain the complete eigen spectrum in all the $M_S$ sectors; this 
is used to compute the magnetic susceptibility of the system. 
As the magnetic measurements are carried out under an applied magnetic 
field, we include a Zeeman term in our calculation which 
contributes an energy $-g\mu_B H_z M_S$ to the eigenstates in a given $M_S$ sector; 
$g$ is the gyromagnetic ratio, $\mu_B$ is the Bohr magneton and 
$H_z$ is the applied magnetic field. The magnetic 
susceptibility of the system is given by,

\begin{equation}
\label{eqn-chiT}
\chi T=\frac{N_Ag^2\mu_B^2F(J,T)}{k_B}
\end{equation}

\begin{equation}
\label{eqn-chiT-FJT}
F(J,T) = \langle M_S^2 \rangle = \frac{\sum_S\sum_{M_S} M_S^2e^{-\frac{E_o(S,M_S)}{k_BT}}}{\sum_S\sum_{M_S} e^{-\frac{E_o(S,M_S)}{k_BT}}}
\end{equation}

In the above expression $N_A$ is the Avogadro number, $k_B$ is the 
Boltzmann’s constant and $E_o (S,M_S)$ are energies of the 
unperturbed Hamiltonian corresponding to the eigen state with 
z-component of total spin $M_S$ \cite{Kahn-Mol Mag}. We also add a Curie contribution 
(C) to the total susceptibility to account for any unreacted 
residual spin moments left after the synthesis. 
Besides, our magnetic data shows that the high temperature 
susceptibility is larger than the 0.0076 emu K/(g Oe) expected 
for free spin-5/2 moments. The $\chi T$ value also shows a small linear 
increase with temperature, contrary to the temperature independent behaviour expected 
in the paramagnetic region for a Curie paramagnet. 
This suggests that there is an additional temperature 
independent susceptibility term or the van Vleck paramagnetic ($\chi_{VV}$)
contribution coming from the excited states. 
The total $\chi T$ value is given by $\chi T= \chi T(ex)+ \chi T(res)+ \chi_{VV} T$.

The strength of various exchange interactions are 
obtained from the parameters that best fit the 
experimental magnetic data. The experimental magnetic data is fitted in 
the temperature range 2-300 K (Figure \ref{figure10}c) and best fit 
parameters correspond to $J_1$= -3.6 K, $J_2$ = -0.94 K, 
$J_3$ = -0.76 K, $g$ = 2.01,  $C$ = 9 x 10$^{-5}$ emu K /(g Oe) and 
$\chi_{VV}$ = 8 x 10$^{-7}$ emu / (g Oe). The contribution to the susceptibilty 
from Curie like and temperature independent paramagnetic 
concentrations are less than the 
3\% of the paramagnetic susceptibility of the system obtained by 
turning off all the exchange interactions. The ground state 
of the system is a spin singlet ($S_{GS}$=0) (Figure \ref{figure10}b), 
confirming an overall 
antiferromagnetic interaction, as also evidenced from the 
decreasing $\chi T$ value as we approach zero Kelvin. The 
first excited state is an $S$=5 state (spin is scaled 
from s = 1/2 to s = 5/2) with an 
energy gap from the ground state of 0.74 K (Figure \ref{figure10}b). 
Besides, there are two more S = 5 states at 3.44 K and 3.54 K, 
before an excited singlet state is found at 3.68 K.
Application of magnetic field can significantly lower the energies 
of states with non-zero magnetization belonging to this S = 5 multiplet. This can lead to 
trapping of moments in higher magnetization states when the system 
is cooled under the influence of magnetic field, resulting in the 
bifurcation of ZFC and FC curves. This is more dominant at low field 
strengths as the population of the high spin state is not saturated 
at these field strengths. Thus, for a small applied field, one observes 
a substantial change in magnetization on cooling. However, at high field 
strengths, at about 3 K, the high spin population is almost saturated and 
this will lead to smaller  change in magnetization as the system is cooled. 
Hence the ZFC and FC susceptibility curves lie very close to each other. 
The first excited state with spin S = 5 at 0.74 K, has significant thermal 
population when cooled to 1.7 K which is the lowest temperature at which 
the study is carried out.  The expectation values of the site $s_z$ operator 
in this S = 5 excited state spin manifold corresponding to $M_S$ = +5 for the best fit exchange 
parameter values are presented in the Figure \ref{figure10}a. Our computed 
spin densities are consistent with the magnetic structure obtained from the neutron 
diffraction measurement, which shows parallel spin arrangement 
along the $a$ and $c$ directions, while the moments are 
anti-parallel along the \{011\} direction.

\section{CONCLUSIONS}
In summary, the measurement of  magnetic properties of a 
Vanthoffite mineral $Na_6Mn(SO_4)_4$ shows antiferromagnetic 
characteristics below 3 K. Neutron diffraction refinements 
at 1.7 K clearly shows an antiferromagentic spin arrangement 
in $bc$-plane of the structure. Numerical results from full 
diagonalization approach support the experimental 
results and unambiguously show the presence of antiferromagntic 
interactions and singlet magnetic ground state in $Na_6Mn(SO_4)_4$.

\section{ACKNOWLEDGEMENTS}
T.N.G. and S.R. would like to thank DST and INSA for financial support. A.D. would like to thank IISc for funding.




\end{document}